\documentstyle[11pt]{article}
\newcommand{\be}{\begin{equation}}
\newcommand{\ee}{\end{equation}}
\newcommand{\ba}{\begin{eqnarray}}
\newcommand{\ea}{\end{eqnarray}}
\newcommand{\pa}{\partial}
\newcommand{\al}{\alpha}
\begin{document}
\title{\bf Schwarz-Sen duality made fully local}
\author{R.L.~Pakman\thanks{Becario CONICET}
\\
{\normalsize\it Departamento de F\'\i sica, Universidad Nacional de La Plata}\\
{\normalsize\it
C.C. 67, 1900 La Plata, Argentina}}
\maketitle
\begin{abstract}
\noindent
Duality symmetric electromagnetic action \`a la Schwarz-Sen is shown
to appear naturally in a chain of equivalent actions which interchange
equations of motion with Bianchi identities. Full symmetry of the
electromagnetic stress tensor is exploited by generalizing this
duality symmetric action to allow for a space-time dependent
mixing angle between electric and magnetic fields. The rotated
fields are shown to satisfy Maxwell-like equations which involve
the mixing angle as a parameter, and a generalized gauge invariance
of the new action is established.
\end{abstract}
\newpage
\section*{Introduction}
Current interest in duality in string theory has brought about a wealth
of studies on  similar symmetries present in other contexts,
such as abelian p-forms theories. The simplest among the latter,
namely free electromagnetism, has long been known to remain
invariant under the interchange of equations of motion
with Bianchi
identities. The first attempt to implement this symmetry at the level of the
basic fields of the action \cite{DT} involved non local transformation among
the $A$ and $E$ fields, in the Hamiltonian (first order) version of the
electromagnetic action. Later on, a non covariant action
was proposed by Schwarz and Sen \cite{SS}
where the transformation is made local at the expense
of doubling the number of gauge fields. When the equations of motion for some
of the fields are used, the usual Maxwell action is recovered.

As it is well known, duality symmetry is actually more general than the discrete interchange of electric
with magnetic fields. It is a continuous symmetry, which gets reduced to a
$U(1)$ group when invariance of the symmetric stress tensor\footnote{All future references to the stress
tensor will be to the symmetric gauge-invariant one.}
is also imposed
\cite{DGHT}. The conserved momentum
associated with this continuous symmetry is found to be the integral of
Chern-Simons terms, and full equivalence between Maxwell and Schwarz-Sen
actions has been shown to remain valid at the quantum level \cite{GIROTTI}.
Moreover, covariant generalizations of the Schwarz-Sen action have been found
by introducing either an infinite number of auxiliary gauge fields \cite{YU}
or finite additional fields in a non polynomical way \cite{PST}.

On the other hand, when considered as a symmetry of the stress tensor, this duality symmetry is
not the most general invariance, because the stress tensor is also
invariant under rotations with a different angle at each spacetime point.
We call this symmetry of the stress tensor fully-local duality, to
distinguish it from that of the Schwarz-Sen action,
which is usually called local duality (in opposition to the non-local transformations
in \cite{DT}). In fact, from our perspective, the duality symmetry of the Schwarz-Sen action could be called global. Our work is
reminiscent of the passage
from global to local gauge invariance of complex matter fields, wherein
gauge fields are introduced in the covariant derivative.

The purpose of this work is twofold. We first show how the Schwarz-Sen action emerges
from a chain of equivalent actions which interchange
equations of motion with Bianchi identities.
Then we rewrite the Schwarz-Sen action in terms of a complex gauge field.
We proceed afterwards to probe the effects of such space-time
dependent rotations on Maxwell
equations and their implementation in a generalized version of the
Schwarz-Sen action. We find that gauge invariance remains valid when suitable generalized.
At the end we present the conclusions.

\section*{The Schwarz-Sen action}

Free electromagnetism equations of motion\footnote{We use the metric $g_{\mu \nu} = diag (+1,-1,-1,-1)$.},
\be
\pa_\mu F^{\mu\nu}[A_\mu]=0 \,,
\ee
with
\be
\label{11}
F_{\mu\nu}[A_\mu] = \pa_\mu A_\nu - \pa_\nu A_\mu \,\,,
\ee
can be obtained from the action
\be
\label{10}
S[A_\mu ]= -\frac{1}{4} \int d^4x F^{\mu\nu}F_{\mu\nu} \,\,,
\ee
and Bianchi identities
\be
\pa_\mu \tilde{F}^{\mu\nu}[A_\mu]=0\,,
\ee
hold automatically, with
\be
\label{12}
\tilde{F}^{\mu\nu}[A_\mu]=\frac{1}{2}\epsilon^{\mu\nu\rho\sigma}F_{\rho\sigma}[A_\mu]\,\,.
\ee
Alternatively, we can vary
\be
\label{13}
S[Z_\mu ]= -\frac{1}{4} \int d^4x \tilde{F}^{\mu\nu} \tilde{F}_{\mu\nu}
\ee
with
\be
\label{14}
\tilde{F}_{\mu\nu}[Z_\mu] = \pa_\mu Z_\nu - \pa_\nu Z_\mu.
\ee
obtaining
\be
\pa_\mu \tilde{F}^{\mu\nu}[Z_\mu]=0
\ee
as equations of motion and
\be
\pa_\mu F^{\mu\nu}[Z_\mu]=0
\ee
as Bianchi identities, with
$F^{\mu\nu}[Z_\mu]$ defined by a relation analogous to Eq.(\ref{12}).

If we define
$E^i=F^{i0}$ and $B^i= -\frac{1}{2}\epsilon^{ijk}F_{jk}$\,, regardless of whether $F_{\mu \nu}$ depends on $A_\mu$ or $Z_\mu$, we see that the effect of passing from $S[A_\mu]$ to $S[Z_\mu]$ is to interchange equations of motion with Bianchi identities.

Now, $S[A_\mu ]$ is equivalent to the first order action
\ba
\label{15}
S[A_\mu , F_{\mu\nu}]&=& \int d^4x
\left[
\frac{1}{4} F^{\mu\nu}F_{\mu\nu}
- \frac{1}{2} F_{\mu\nu}(\pa^\mu A^\nu - \pa^\nu A^\mu)
\right]\\
&=& \int d^4x
\left[
\frac{1}{2} \left(\bf{B}^2 - \bf{E}^2 \right)
-\bf{E \cdot \dot{A}} -\bf{E \cdot \nabla}A_0
-\bf{B \cdot \nabla \times A}
\right]
\nonumber
\ea
where
$A^{\mu} = (A^0,\bf{A})$.
As it is well known, $A_\mu $ and $F_{\mu\nu}$ are independent fields in this approach, but varying with respect to $F_{\mu\nu}$ definition (\ref{11}) is recovered, and replacing it in $S[A_\mu , F_{\mu\nu}]$, we get back
$S[A_\mu ]$. Now, varying $S[A_\mu , F_{\mu\nu}]$ with respect to $B^i$ and $A_0$, yields $\bf{B}=\bf{\nabla \times A}$ and
$\bf{\nabla \cdot E}=0$, respectively. Replacing these last equations in (\ref{15}) we get
\be
S[A_i, E_T^i]= \int d^4x
\left\{\
 E_T^i \dot{A}_i
-\frac{1}{2} \left[
{\bf E_T^2} + \left( {\bf \nabla \times A} \right)^2
\right]
\right\}\
\ee
where $\bf{E_T}$ indicates that only the transversal part of $\bf{E}$ survives.
Hence, another vector potential can be introduced through
$\bf{E_T} = \bf{\nabla \times Z}$. As we shall see, ${\bf Z}$ will be later identified with the spatial components of  the tetravector $Z_\mu$, already introduced.  Then $S[A_i, E_T^i]$ can be written as

\be
S[A_i,Z_i] = -\frac{1}{2} \int d^4x
\left[
\bf{\nabla \times Z \cdot \dot{A}}
- \bf{\nabla \times A \cdot \dot{Z}}
+\left( \bf{\nabla \times A} \right)^2
+\left( \bf{\nabla \times Z} \right)^2
\right] \,,
\ee
where an integration by parts has been performed, thus exhibiting the symmetry between $\bf{Z}$ and $\bf{A}$. The equations of motion are now
\ba
\label{19}
\bf{\dot{E}} &=& \bf{\nabla \times \dot{Z}} = \bf{ \nabla \times \nabla \times A} = \bf{\nabla \times B}
\\
\bf{\dot{B}} &=& \bf{\nabla \times \dot{A}} = - \bf{ \nabla \times \nabla \times Z} = -\bf{\nabla \times E}
\nonumber
\ea
Some comments are in order.
First note that the assignment in $ S[A_i, Z_i ] $ of (\ref{19}) as equations of motion and ${\bf \nabla \cdot E} = {\bf \nabla \cdot B}=0$ as ``Bianchi identities'', corresponds neither to $S[A_\mu]$ nor to $S[Z_\mu]$. It is a mixture between the assignments in both actions.
Secondly, we could go through steps similar to those that led us from $S[A_\mu]$ to $S[A_i,Z_i]$, but now in reverse order and  with $Z_\mu$ taking the place of $A_\mu$. The antisymmetric disposition of $\bf{Z}$ and $\bf{A}$ in the first terms of $S[A_i,Z_i]$ necessitates the definition
$Z_\mu = (Z_0,\bf{Z})$, which should be compared to
$A^{\mu} = (A^0,\bf{A})$.
We would end up with an action which can be identified with $S[Z_\mu]$ as defined in (\ref{13}).

At this point, Schwarz-Sen action is obtained from $S[A_i,Z_i]$ noting that
for any functions $A_0$ and $Z_0$,
\be
\label{18}
\int d^4x {\bf \nabla}  A_0 {\bf \cdot \nabla \times Z} =
\int d^4x {\bf \nabla}  Z_0 {\bf \cdot \nabla \times A} =0\,\,,
\nonumber
\ee
so that
\ba
\label{99}
S[A_i,Z_i] &=& S[A_\mu ,Z_\mu ] =
\nonumber
\\
&=& -\frac{1}{2} \int d^4x
\left[
{\bf\nabla \times Z \cdot} \left( {\bf \dot{A}} + {\bf\nabla}A_0 \right)
- {\bf\nabla \times A \cdot} \left( {\bf \dot{Z}} + {\bf\nabla}Z_0 \right)
\right.
\nonumber
\\
&&\left.
+\left( {\bf \nabla \times A} \right)^2
+\left( {\bf \nabla \times Z} \right)^2
\right] \,,
\ea
which is easily recognized as the Schwarz-Sen action.
Eq.(\ref{99}) summarizes one of the main results of this work:
we see that $S[A_\mu,Z_\mu]$ emerges in the middle point of a chain of equivalent actions that lead from $S[A_\mu]$ to $S[Z_\mu]$ and backwards, interchanging equations of motion with Bianchi identities.

It should be stressed that the $A_0$ and $Z_0$ fields of the Schwarz-Sen action bear no relation to those which appear in $S[A_\mu]$ and  $S[Z_\mu]$. In the former case, advantage is taken of relation (\ref{18}) to
make the integral nicely dependent on two four-potentials, while in the latter case they are used to impose the constraints ${\bf \nabla \cdot E} = 0$ and ${\bf \nabla \cdot B}=0$  .

\section*{Complex field formulation}

If we define ${\bf  \Phi }= {\bf A} + i{\bf  Z}$ \,,  $\Phi_0 = A^0 + iZ_0$ \,, $\Phi_\mu = (\Phi_0, { \bf  \Phi})$
\footnote{With our choice $\Phi_\mu $ is not a tetravector since in its definition covariant and contravariant components are summed. Other choices would render it a tetravector.},
Schwarz-Sen action can be written as

\ba
S[\Phi_\mu, \Phi_ {\mu *}] &=&  - \frac{1}{2} \int d^4x
\left\{\
\frac{i}{2}
\left[
\left( {\bf \dot{\Phi}} + {\bf \nabla}\Phi_0 \right)
\cdot {\bf\nabla \times \Phi}^*
\right.
\right.
\\
\nonumber
&&-
\left.
\left.
\left( {\bf \dot{\Phi}^*} + {\bf \nabla}\Phi_0^* \right)
\cdot {\bf \nabla \times \Phi}
\right]
+
{\bf \nabla \times \Phi \cdot \nabla \times \Phi}^*
\right\}\
\,\,.
\ea
Varying with respect to $\Phi^{\mu*}$ yields
\be
{\bf \nabla \times \dot{\Phi}} = i{\bf \nabla \times \nabla \times \Phi}
\ee
which is the same as Eq.(\ref{19}), while variation with respect to $\Phi^\mu$ yields the complex conjugate equation.

$ S[\Phi^\mu, \Phi^{\mu *}]$ is separately invariant under the local gauge transformations
\ba
\label{31}
&& {\bf  \Phi}  \rightarrow  {\bf  \Phi} + {\bf  \nabla} \Psi_1
\\
&& {\bf  \Phi^*}  \rightarrow  {\bf  \Phi^*} + {\bf  \nabla} \Psi_2
\label{32}
\\
&& \Phi_0  \rightarrow  \Phi_0 + \Xi_1
\\
&& \Phi_0^*  \rightarrow  \Phi_0^* + \Xi_2
\ea
where $\Psi_1$, $\Psi_2$, $\Xi_1$ and $\Xi_2$ are arbitrary gauge functions satisfying apropiate boundary conditions. In case we want the surface term picked by the Lagrangian to be real, conditions $\Psi_1 = \Psi_2^* $ and $\Xi_1 = \Xi_2^*$ should be further imposed. The Lagrangian is also invariant under the simultaneous global U(1) duality rotations
\ba
\label{33}
&&\Phi_\mu  \rightarrow e^{i \al} \Phi_\mu
\\
&&\Phi_\mu^*  \rightarrow e^{-i \al} \Phi_\mu^*.
\ea
Under rotation (\ref{33}),
\be
{\bf  F} \equiv {\bf \nabla \times \Phi} = {\bf B} + i {\bf E}
\ee
is transformed to
\be
e^{i \al}{\bf F} = (\cos \al {\bf B} - \sin \al {\bf E}) + i (\cos \al {\bf E} + \sin \al {\bf B})\,\,,
\ee
which gets reduced to the known discrete duality transformation for $\al =  \pi /2$. This continuous $U(1)$ symmetry has associated the conserved real current $j^\mu = (j^0, {\bf j})$
\footnote{Note that $j^\mu$ is not a tetravector, since the action $ S[\Phi_\mu, \Phi_ {\mu }^*]$ is not a scalar.}
, where
\ba
&&j^0 =  \frac{1}{4} \left(
{\bf  \nabla \times \Phi^* \cdot \Phi} + {\bf  \nabla \times \Phi \cdot \Phi^*}
\right)
\\
\nonumber
&&{\bf j} = \frac {1}{4} {\bf \nabla \times} ( \Phi_0 {\bf \Phi^*} +  \Phi_0^* {\bf \Phi} )
+ \frac {1}{4} ( {\bf  \dot{\Phi}^* \times \Phi} +  {\bf  \dot{\Phi} \times \Phi}^*)
\\
\nonumber
&&+ \frac {i}{2} \left[
{\bf  (\nabla \times \Phi^*) \times \Phi} - {\bf  (\nabla \times \Phi) \times \Phi^*}
\right] \,\,.
\ea
In terms of the original ${\bf A}$ and ${\bf Z}$ fields, the conserved momentum reads
\be
\int d^3x\,\, {\bf \nabla \times A \cdot A} + {\bf \nabla \times Z \cdot Z}
\ee
which is of the usual Chern-Simons type.

\section*{Fully-local duality}
Given a configuration of electromagnetic fields
\be
{\bf  F'}  = {\bf B'} + i {\bf E'}
\ee
which satisfies Maxwell equations
\ba
&&{\bf \nabla \cdot F'}=0
\\
\nonumber
&& {\bf  \dot{F}'}= i{\bf \nabla \times F'}\,\,,
\ea
if we rotate them through
\be
\label{40}
{\bf F}= e^{i \al (x)} {\bf F'}\,\,,
\ee
with a space-time dependent angle $\al (x)$, the rotated fields satisfy
\ba
\label{41}
&&{\bf D \cdot F}=0
\\
\label{42}
&& D_t{\bf F}= i{\bf  D \times F}\,\,,
\ea
where
\be
D_\mu = (\pa_\mu -i \pa_\mu \al) \ee is a kind of covariant
derivative. Now, taking the real and imaginary parts of
Eqs.(\ref{41}-\ref{42}) we have a total of eight equations. Were
one to solve the four $\pa_\mu \al$ therefrom, they would be
overdetermined. The four compatibility conditions turn out to be
\ba \label{43} &&\frac{1}{2} \pa_t ( {\bf E^2} + {\bf B^2}) + {\bf
\nabla \cdot (E \times B) } = 0
\\
\nonumber
&&\pa_t ({\bf E \times B}) + {\bf E \times \nabla \times E} + {\bf B \times \nabla \times B}
-{\bf  B(\nabla \cdot B)} - {\bf  E(\nabla \cdot E)} = 0
\ea
which are just the conservation equations of the stress tensor, but now
for the rotated fields. This is actually not a piece of  news, since the stress tensor is
invariant under (\ref{40}). Of course, not every configuration satisfying the
four Eqs.(\ref{43}) comes from a U(1)-local rotation of another configuration
which satisfies Maxwell equations, since the $\pa_\mu \al$ field solved
from Eqs.(\ref{41}-\ref{42}) must satisfy
the integrability conditions $\pa_\mu \pa_\nu \al =\pa_\nu \pa_\mu \al $.

Now, since the original fields ${\bf F'}$ were derived from
\be
{\bf F'}= {\bf \nabla \times \Phi'}\,\,,
\ee
the rotated fields can be derived from a rotated potential using a covariant rotor
\ba
{\bf F} &=& e^{i \al (x)}{\bf F'}={\bf D \times \Phi}
\\
\nonumber
{\bf \Phi} &=& e^{i \al (x)}{\bf  \Phi'}
\ea
Eq.(\ref{42}) reads now
\be
\label{44}
D_t{\bf D \times \Phi} = i{\bf D \times D \times \Phi}
\ee
This equation can be obtainted from the following generalization of the Schwarz-Sen action,
\ba
S[\Phi_\mu, \Phi_ {\mu *}, \al] &=&  - \frac{1}{2} \int d^4x
\left\{\
\frac{i}{2}
\left[
\left( D_t{\bf  \Phi} + {\bf D}\Phi_0 \right)
\cdot {\bf D \times \Phi}^*
\right.
\right.
\\
\nonumber
&&-
\left.
\left.
\left(  D_t{\bf \Phi^*} + {\bf D}\Phi_0^* \right)
\cdot {\bf D \times \Phi}
\right]
+
{\bf D \times \Phi \cdot D \times \Phi}^*
\right\}\
\,\,.
\ea
Variation with respect to $\Phi_\mu$ yields Eq.(\ref{44}), and with respect to $\Phi_\mu^*$ yields it complex conjugate. Varying with respect to $\al$ yields no new equations.

$ S[\Phi^\mu, \Phi^{\mu *},\al]$ is now separately invariant under the local transformations
\ba
&& {\bf  \Phi}  \rightarrow  {\bf  \Phi} + {\bf  D} \Psi_1
\\
&& {\bf  \Phi^*}  \rightarrow  {\bf  \Phi^*} + {\bf  D^*} \Psi_2
\\
&& \Phi_0  \rightarrow  \Phi_0 + \Xi_1
\\
&& \Phi_0^*  \rightarrow  \Phi_0^* + \Xi_2
\ea
where the conditions $\Psi_1 = \Psi_2^* $ and $\Xi_1 = \Xi_2^*$ should again be imposed in case we want the surface term picked by the Lagrangian to be real. The Lagrangian is also invariant under the simultaneous local U(1) rotations
\ba
\label{45}
&&\Phi_\mu  \rightarrow e^{i \beta (x)} \Phi_\mu
\\
&&\Phi_\mu^*  \rightarrow e^{-i \beta(x)} \Phi_\mu^*\,\,,
\ea
together with
\be
\label{46}
\al \rightarrow \al + \beta\,\,.
\ee
Under the transformations (\ref {45}-\ref{46}), the fields
\be
{\bf  F} =  {\bf D \times \Phi} = {\bf B} + i {\bf E}
\ee
are rotated into
\be
e^{i \beta(x)}{\bf F} = (\cos \beta(x) {\bf B} - \sin \beta(x) {\bf E}) +
 i (\cos \beta(x) {\bf E} + \sin \beta(x) {\bf B})\,\,.
\ee
We see that the action $S[\Phi_\mu, \Phi_ {\mu *}]$ is a particular case
of  $S[\Phi_\mu, \Phi_ {\mu *}, \al]$ , which is obtained from the latter through the transformations
(\ref {45}-\ref{46}) with $\beta = -\al$.

The conserved current associated to the $U(1)$ symmetry in the generalized action is now
\ba
&&j^0 =  \frac{1}{4} \left(
{\bf  D \times \Phi^* \cdot \Phi} + {\bf  D \times \Phi \cdot \Phi^*}
\right)
\\
\nonumber
&&{\bf j} = \frac {1}{4} {\bf \nabla \times} ( \Phi_0 {\bf \Phi^*} +  \Phi_0^* {\bf \Phi} )
+ \frac {1}{4} ( {\bf  D_t {\Phi}^* \times \Phi} +  {\bf  D_t{\Phi} \times \Phi}^*)
\\
\nonumber
&&+ \frac {i}{2} \left[
{\bf  (D \times \Phi^*) \times \Phi} - {\bf  (D \times \Phi) \times \Phi^*}
\right] \,\,.
\ea

\section*{Conclusions}
We have shown that the duality symmetric Schwarz-Sen action is the middle point of a chain of equivalent actions, which interchange equations of motion with Bianchi identities. Space-time dependent duality rotations were studied and the equations obeyed by the rotated fields were obtained, along with a generalized action from which these equations are derived. Remarkably, the gauge symmetries of this action are a natural extension of those of the Schwarz-Sen action.

An important property of the new equations (\ref{41}-\ref{42}) is that for any $\al(x)$, every solution of them is mapped one-to-one to a solution of Maxwell equations. Among further developments of this work would be to show how this equivalence holds at the quantum level. This involves an analysis of the constrain structure of $S[\Phi_\mu, \Phi_ {\mu *}, \al]$. Moreover, coupling to external currents in a fully-local-duality-preserving way is also worth studying.  We hope to deal with these topics in a next work.

\section*{Acknowledgements}
We are indebted to Nicol\'as Grandi and Enrique Moreno for useful
comments. We also wish to thank Fidel Schaposnik for constant
encouragement during this work.

\end{document}